\documentclass[aps,prb,twocolumn,superscriptaddress]{revtex4}
\usepackage{amsmath,amssymb}
\usepackage{graphicx}
\usepackage{times}

\begin{document}
\title{Evolution of magnetic, transport, and thermal properties in Na$_{4-x}$Ir$_3$O$_8$}
\author{Ashiwini Balodhi}
\affiliation{Indian Institute of Science Education and Research (IISER) Mohali, Knowledge city, Sector 81, Mohali 140306, India}
\author{A. Thamizhavel}
\affiliation{Tata Institute of Fundamental Research, Homi Bhabha Road, Colaba, Mumbai 400 005, India}
\author{Yogesh Singh}
\affiliation{Indian Institute of Science Education and Research (IISER) Mohali, Knowledge city, Sector 81, Mohali 140306, India}
\date{\today}
 
\begin{abstract}
The hyper-kagome material Na$_4$Ir$_3$O$_8$ is a three-dimensional spin-liquid candidate proximate to a quantum critical point (QCP).  We present a comprehensive study of the structure, magnetic susceptibility $\chi$, heat capacity $C$, and electrical transport on polycrystalline samples of the doped hyper-kagome material Na$_{4-x}$Ir$_3$O$_8$ ($x \approx 0, 0.1, 0.3, 0.7)$.  Materials with $x\leq 0.3$ are found to be Mott insulators with strong antiferromagnetic interactions and no magnetic ordering down to $T = 2$~K\@.  All samples show irreversibility below $T \approx 6$~K between the zero-field-cooled and field-cooled magnetization measured in low fields ($H = 0.050$~T) suggesting a frozen low temperature state although no corresponding anomaly is seen in the heat capacity.  The $x = 0.7$ sample shows $\rho(T)$ which weakly increases with decreasing temperature $T$, nearly $T$ independent $\chi$, a linear in $T$ contribution to the low temperature $C$, and a Wilson ratio $R_W \approx 7$ suggesting anomalous semi-metallic behavior.

\end{abstract}

\maketitle
Intense research on geometrically frustrated magnets has led to a plethora of exciting new physics in the recent past (see \onlinecite{Mila2000, Ramirez2008, Balents2010} for recent reviews).  The spin-ice phase in pyrochlore magnets \cite{Harris1997, Bramwell2001}, quantum spin-liquid (QSL) in triangular lattice organic compounds $\kappa$-(BEDT-TTF)$_2$Cu$_2$(CN)$_3$ \cite{Yamashita2008, Yamashita2009} and EtMe$_3$Sb[Pd(dmit)$_2$]$_2$ \cite{Yamashita2010}, and in 2D kagome lattice inorganic materials ZnCu$_3$(OH)$_6$Cl$_2$, \cite{Helton2007, Han2012} and BaCu$_3$V$_2$O$_8$(OH)$_2$ \cite{Okamoto2009}, the possibility of magnetic monopoles as excitations of the spin-ice state \cite{Castelnovo2008, Morris2009, Fennell2009, Bramwell2009}, and heavy fermion behavior in the itinerant frustrated magnet LiV$_2$O$_4$ \cite{Kondo1997} are just a few examples.  

Na$_4$Ir$_3$O$_8$, with effective spins $S = {1\over 2}$ on a frustrated hyperkagome lattice may be the first candidate QSL with a 3-dimensional (3D) structure \cite{Okamoto2007, Singh2013, Hopkinson2007,Lawler2008a, Zhou2008,Lawler2008b, Podolsky2009}.  Heat capacity measurements down to $T = 0.5$~K have shown an absence of long-range magnetic ordering \cite{Singh2013} even though magnetic susceptibility provides evidence for strong antiferromagnetic interactions as evidenced by a Weiss temperature of $\theta \sim - 600$~K~~ \cite{Okamoto2007, Singh2013}.  Several anomalous features , apparently associated with the spin-liquid state, have been observed in thermodynamic measurements on Na$_4$Ir$_3$O$_8$.  These include an anomaly around $T \sim 30$~K in the magnetic heat capacity, a $T^n$ dependence of the heat capacity at low temperatures with an exponent $n$ between $2$ and $3$, and a large Wilson ratio $R_W \approx 30~~$ \cite{Okamoto2007, Singh2013}.  Bulk susceptibility measurements have shown signs of spin freezing at low temperastures ($T \sim 5$~K) which were suggested to be arising from a small fraction of the sample \cite{Okamoto2007}.  Recently microscopic muSR and neutron scattering measurements down to $T = 20$~mK have shown absence of long-range magnetic order on polycrystalline samples.  However, these measurements reveal that bulk Na$_4$Ir$_3$O$_8$ goes into a short-range frozen state with quasi-static moments below $T = 6$~K which is either disorder driven or is stabilized by quantum fluctuations~ \cite{Dally2014}.  It is thus unclear what role disorder plays in stablizing this frozen state and whether better/ideal samples would show spin-liquid behavior down to $T \rightarrow0$~K\@.  Nevertheless the frozen state occurs at a temperature ($T \sim 5$~K) which is two orders of magnitude smaller than the Weiss scale classifying Na$_4$Ir$_3$O$_8$ as a highly frustrated magnet.  

Introducing charge carriers into a geometrically frustrated Mott insulating state is expected to lead to anomalous metallic properties and even unconventional superconductivity.  The high-$T_c$ cuprates and the organic Mott insulators are such examples where carrier doping and/or modest pressures can tune the system from antiferromagnetic or spin-liquid insulators to superconductors (see \onlinecite{Johnston, Powell2011} for reviews).  It is known that the charge gap in Na$_4$Ir$_3$O$_8$ is small ($\sim 500$ -- $1000$~K) and that it is close to a metal-insulator transition \cite{Okamoto2007, Singh2013}.  There is also recent experimental evidence that Na$_4$Ir$_3$O$_8$ maybe situated close to a quantum-critical-point (QCP) \cite{Singh2013}.  

Recently, attempts to hole-dope Na$_4$Ir$_3$O$_8$ have led to the discovery of a new material Na$_3$Ir$_3$O$_8$ having the same hyperkagome Ir sublattice \cite{Takayama2014}.  Na$_3$Ir$_3$O$_8$ is found to be a semi-metal with a small density of states \cite{Takayama2014}.  The thermal conductivity of weakly insulating Na$_4$Ir$_3$O$_8$ was found to be anomalously low compared to semi-metallic Na$_3$Ir$_3$O$_8$ \cite{Fauque2014}.  Na$_3$Ir$_3$O$_8$ however, crystallizes in a different structure to Na$_4$Ir$_3$O$_8$ and one cannot continuously go from one structure to the other \cite{Takayama2014}.  Therefore, what happens on doping the original hyperkagome Na$_4$Ir$_3$O$_8$ structure is still unknown.

To explore the possibility of a nearby metallic (superconducting!) or magnetically ordered state we have tried to tune Na$_4$Ir$_3$O$_8$ away from its Mott insulating state by creating Na deficient samples while keeping the original structure intact.  To this end we have synthesized hole-doped materials Na$_{4-x}$Ir$_3$O$_8$ ($x = 0, 0.1, 0.3, 0.7$) and studied their structural, magnetic and thermal properties.   

Specifically we track the evolution of the Weiss temperature, the anomaly in the magnetic heat capacity, the power-law heat capacity at low temperatures, and the Wilson ratio as increasing amounts of Na are removed from Na$_4$Ir$_3$O$_8$.  We find that the insulating behavior, the local moment magnetism with strong antiferromagnetic interactions persists in the Na deficient samples.  Only in the largest doping $x = 0.7$ do we see a conventional $T^3$ low temperature heat capacity and a much reduced Wilson ratio $R_W \approx 7$.     

\begin{figure}[t]
\includegraphics[width=3 in]{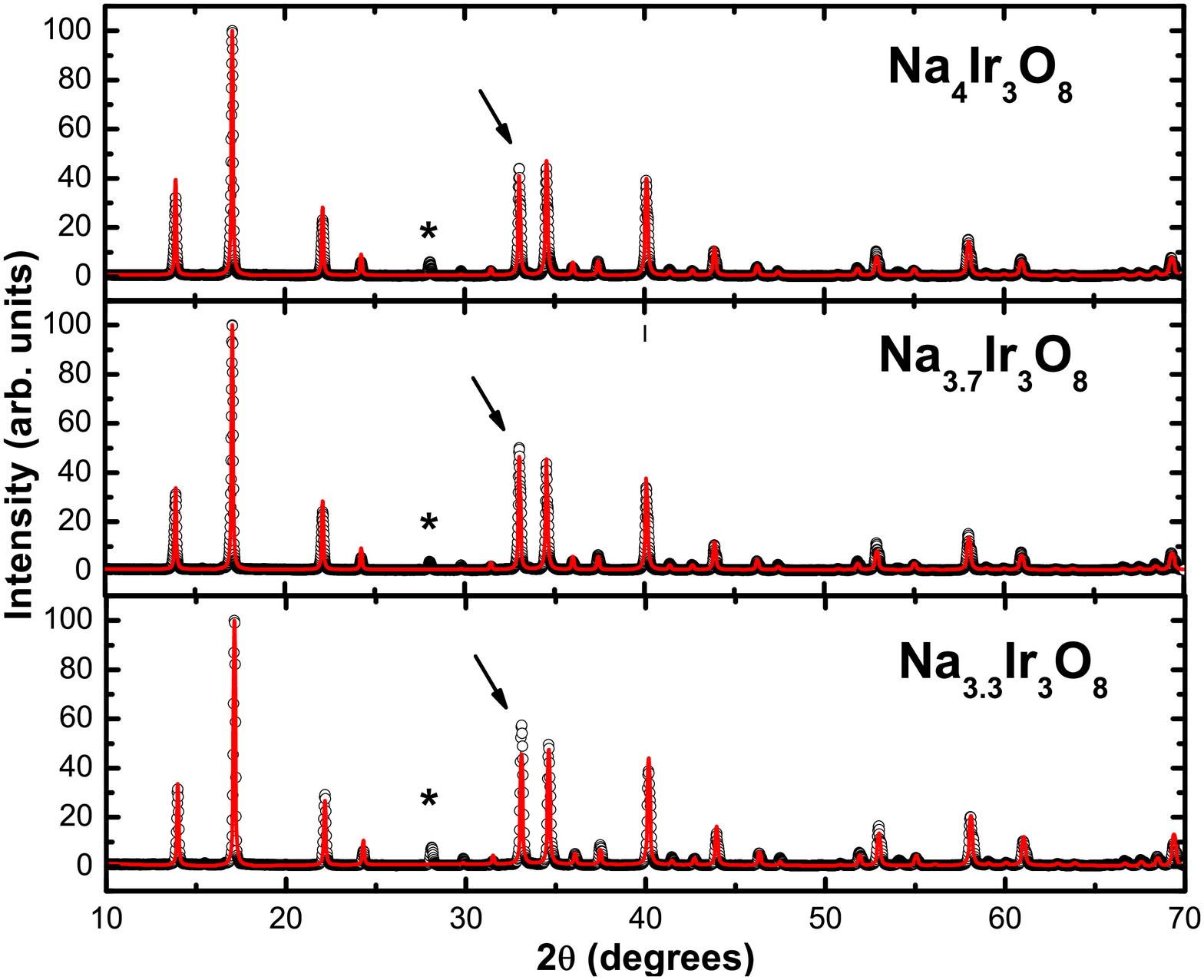}
\caption{(Color online) Powder X-ray diffraction pattern for Na$_{4-x}$Ir$_3$O$_8$ ($x = 0 , 0.3, 0.7$) materials (open symbols).  The solid curve through the data is the Rietveld refinement.  The $*$ marks the position of the largest diffraction peak for the IrO$_2$ impurity phase found in the samples.  The arrow marks the position of the (311) reflection whose intensity increases with Na deficiency.  
\label{Fig-xrd}}
\end{figure}

\begin{table*}

\caption{\label{atomicpositions} Structural parameters obtained from a Rietveld refinemnet of room temperature powder X-ray patterns shown in Fig.~\ref{Fig-xrd} with space group \#213, $P4_132$.  We have used the standardized structural parameters.  The unit cell co-ordinates in this notation are related to the earlier published unit cell by an origin shift of ($1/2, 1/2, 1/2$)}

\begin{ruledtabular}

\begin{tabular}{|c|cccccc|}
Na$_4$Ir$_3$O$_8$ & $a = 8.981(2)$~\AA& & & IrO$_2 \approx 7\%$& & \\ \hline
atom & Wyck &\emph{x} & \emph{y} & \emph{z} & \emph{Occ.} & \emph{B}(\AA)  \\ \hline  

Ir & 12d&1/8 & 0.141(3) & 0.392(5) & 1 & 0.4(1) \\  
Na1 & 4a&3/8 & 3/8 & 3/8 & 1 & 1.8(6) \\ 
Na2 & 4b&7/8 & 7/8 & 7/8 & 0.71(1) & 2.1(4) \\ 
Na3 & 12d&1/8 & 0.883(4) & 0.131(2) & 0.75 & 1.4(8) \\ 
O1 & 8c&0.142(6) & x & x & 1 & 0.9(2) \\ 
O2 & 24e&0.124(6) & 0.315(8) & 0.352(4) & 1 & 1.3(6) \\ 
& & & & & & \\ \hline
Na$_{3.7}$Ir$_3$O$_8$ & $a = 8.979(7)$~\AA& & & IrO$_2 \approx 4\%$& & \\ \hline
Ir & 12d&1/8 & 0.137(3) & 0.383(4) & 1 & 0.8(3) \\  
Na1 & 4a&3/8 & 3/8 & 3/8 & 1 & 2.2(1) \\ 
Na2 & 4b&7/8 & 7/8 & 7/8 & 0.64(1) & 1.7(3) \\ 
Na3 & 12d&1/8 & 0.868(4) & 0.129(4) & 0.70(2) & 2.4(8) \\ 
O1 & 8c&0.105(6) & 0.105(6) & 0.105(6) & 1 & 1.0(7) \\ 
O2 & 24e&0.112(6) & 0.335(8) & 0.337(4) & 1 & 2.1(3) \\ 
& & & & & & \\ \hline
Na$_{3.3}$Ir$_3$O$_8$ & $a = 8.980(3)$~\AA& & & IrO$_2 \approx 6\%$& & \\ \hline
Ir & 12d&1/8 & 0.139(3) & 0.386(5) & 1 & 0.7(2) \\  
Na1 & 4a&3/8 & 3/8 & 3/8 & 1 & 2.1(2) \\ 
Na2 & 4b&7/8 & 7/8 & 7/8 & 0.36(3) & 2.6(4) \\ 
Na3 & 12d&1/8 & 0.876(4) & 0.122(1) & 0.64(2) & 1.9(7) \\ 
O1 & 8c&0.105(6) & 0.105(6) & 0.105(6) & 1 & 0.9(3) \\ 
O2 & 24e&0.112(6) & 0.335(8) & 0.337(4) & 1 & 1.1(4) \\ 
\end{tabular}

\end{ruledtabular}
\label{Table-xrd}
\end{table*} 
Polycrystalline samples of Na$_{4-x}$Ir$_3$O$_8$ ($x = 0 - 0.7$) and Na$_4$Sn$_3$O$_8$ were synthesized using high-purity starting materials Na$_2$CO$_3$ (5N Alfa Aesar) and Ir metal powder (4N, Alfa Aesar) or SnO$_2$ (5N, ALfa Aesar).  Starting materials were mixed in amounts appropriate for a given $x$ and heated in air inside covered alumina crucibles at 750~$^\circ$C for 24~hrs for calcination.  The resulting black powders were thoroughly ground and mixed in an agate mortar and pestle, pressed into pellets, and given two heat treatments at 985~$^\circ$C and 1000~$^\circ$C for 16~hrs each with an intermediate grinding and pelletizing step.  The material was finally quenched in air after the final treatment.  All materials were then stored and handled in an inert gas glove-box (MBraun, Argon, $H_2O < 0.1$~ppm, $O_2 < 0.1$~ppm).  Magnetic susecptibility, heat capacity, and electrical transport measurements down to $T = 2$~K were measured using a Quantum Design PPMS.  The low temperature specific heat for the $x = 0$ sample were measured using the dilution refrigerator option of the PPMS.  

Powder X-ray diffraction patterns for three representative samples Na$_{4-x}$Ir$_3$O$_8$ ($x = 0, 0.3, 0.7$) are shown in Fig.~\ref{Fig-xrd}.  The solid (red) curves through the data are the results obtained from Rietveld refinement of the data.  The parameters obtained from the refinements are listed in Table~\ref{Table-xrd}.  The quoted values of $x$ are those obtained from the refinements.  All samples were found to have the correct majority phase.  The only impurity phase found in the samples was a small amount of IrO$_2$.  The position of the most intense IrO$_2$ reflection is marked with a $*$ in the patterns shown in Fig.~\ref{Fig-xrd}.  The approximate amount of IrO$_2$ in each sample is alo listed in the Table~\ref{Table-xrd}.  From the Table~\ref{Table-xrd} it can be seen that the lattice parameter does not change with Na deficiency.  The major change is in the occupation of the Na2 and Na3 sites which reduces for the Na deficient samples as expected.  Between the two sites, the occupancy of Na2 site reduces more in comparison to Na3.  In the diffraction pattern, the only major change with increasing Na deficiency is the intensity of the $(311)$ reflection which increases relative to the other reflections as indicated by an arrow in the Fig.~\ref{Fig-xrd}.  

Removing Na from Na$_4$Ir$_3$O$_8$ can in principle lead to a two phase mixture of Na$_4$Ir$_3$O$_8$ and Na$_3$Ir$_3$O$_8$.  Both these materials have very similar x-ray diffraction patterns and so lab x-ray diffraction experiments will be unable to distinguish a mixed phase sample from a single phase Na-deficient sample.  The main difference between the structures of these two materials is the number of Na positions.  Na$_4$Ir$_3$O$_8$ has 3 distinct Na positions while Na$_3$Ir$_3$O$_8$ has two.  Only one of the Na positions is common between the two structures.  Therefore, if one were to do a Na NMR measurement at room temperature one would see 3 resonances if only the Na$_4$Ir$_3$O$_8$ structure were present and 4 Na resonances if there was a mixture of Na$_4$Ir$_3$O$_8$ and Na$_3$Ir$_3$O$_8$.  Preliminary Na NMR measurements on our samples have revealed only 3 Na resonances for all samples \cite{Kwang2015}.  We are therefore fairly confident that we are working with Na-deficient Na$_4$Ir$_3$O$_8$ samples.

\begin{figure}[t]
\includegraphics[width=3in]{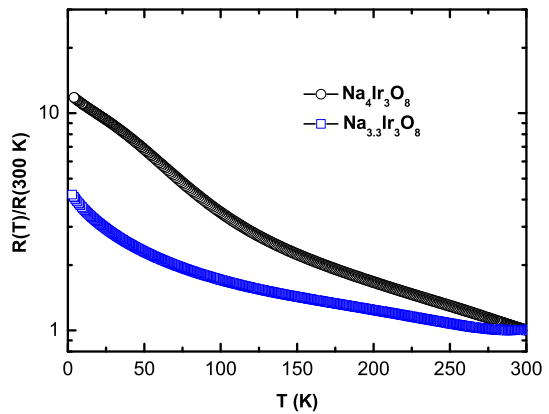}
\caption{(Color online) Resistance divided by the $T = 300$~K value $R(T)/R(300 K)$ versus $T$ of Na$_{4-x}$Ir$_3$O$_8$ ($x = 0, 0.7$).    
\label{Fig-res}}
\end{figure}

The resistance $R(T)$ divided by the $T = 300$~K value is shown in Fig.~\ref{Fig-res} for the $x = 0$ and $x = 0.7$ samples.  The $T = 300$~K resistivity values for the two samples are $41~\rm{m}\Omega~\rm{cm}$ and $16~\rm{m}\Omega~\rm{cm}$, respectively.  Since the samples were pressed sintered pellets, the errors on these absolute values could be up to 50\% due to uncertainity in determining the true dimensions, specially for the $x = 0.7$ sample pellet which were porous.  However, the values and temperature dependence suggest that the conductivity for the $x = 0.7$ sample is slightly enhanced compared to the parent $x = 0$ material.  We note that the amount of IrO$_2$ phase in the $x = 0$ and $x = 0.7$ samples is approximately the same and hence our conclusion made from a comparison of the resistivities of these two samples will not be affaceted by the presence of the impurity phase.  The temperature dependence and room temperature value of the resistivity for our samples are similar to those recently reported for $N_{3.6}$Ir$_3$O$_8$ single crystals \cite{Fauque2014}.  For the $x = 0.7$ sample $R(2\rm{K})/R(300\rm{K}) \approx 3.5$ suggesting that this sample is close to being (semi-)metallic.  Both the magnetic susceptibility $\chi$ and heat capacity $C$ for this sample support this.

\begin{figure}[t]
\includegraphics[width=3.25in]{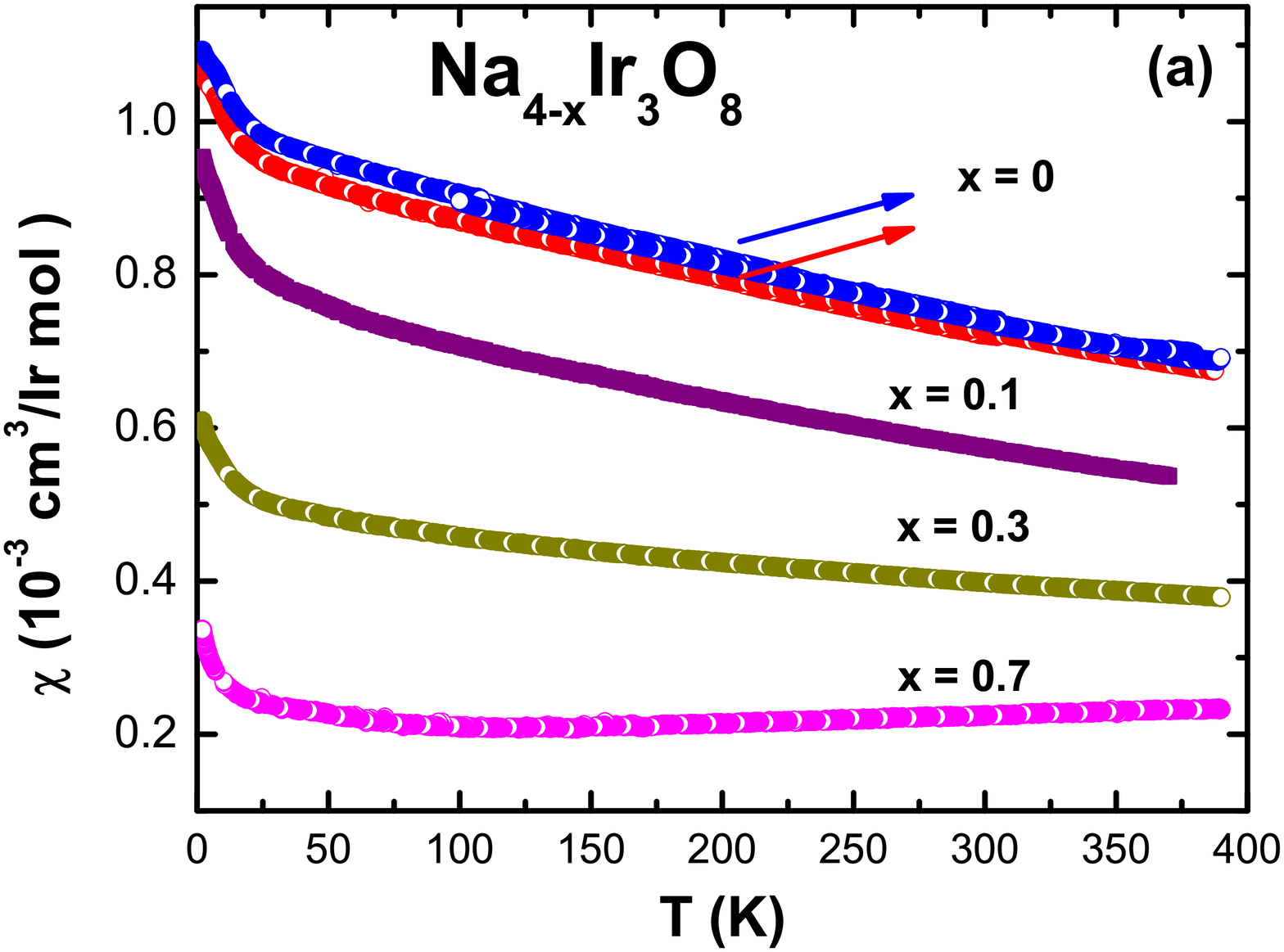}
\includegraphics[width=3.25in]{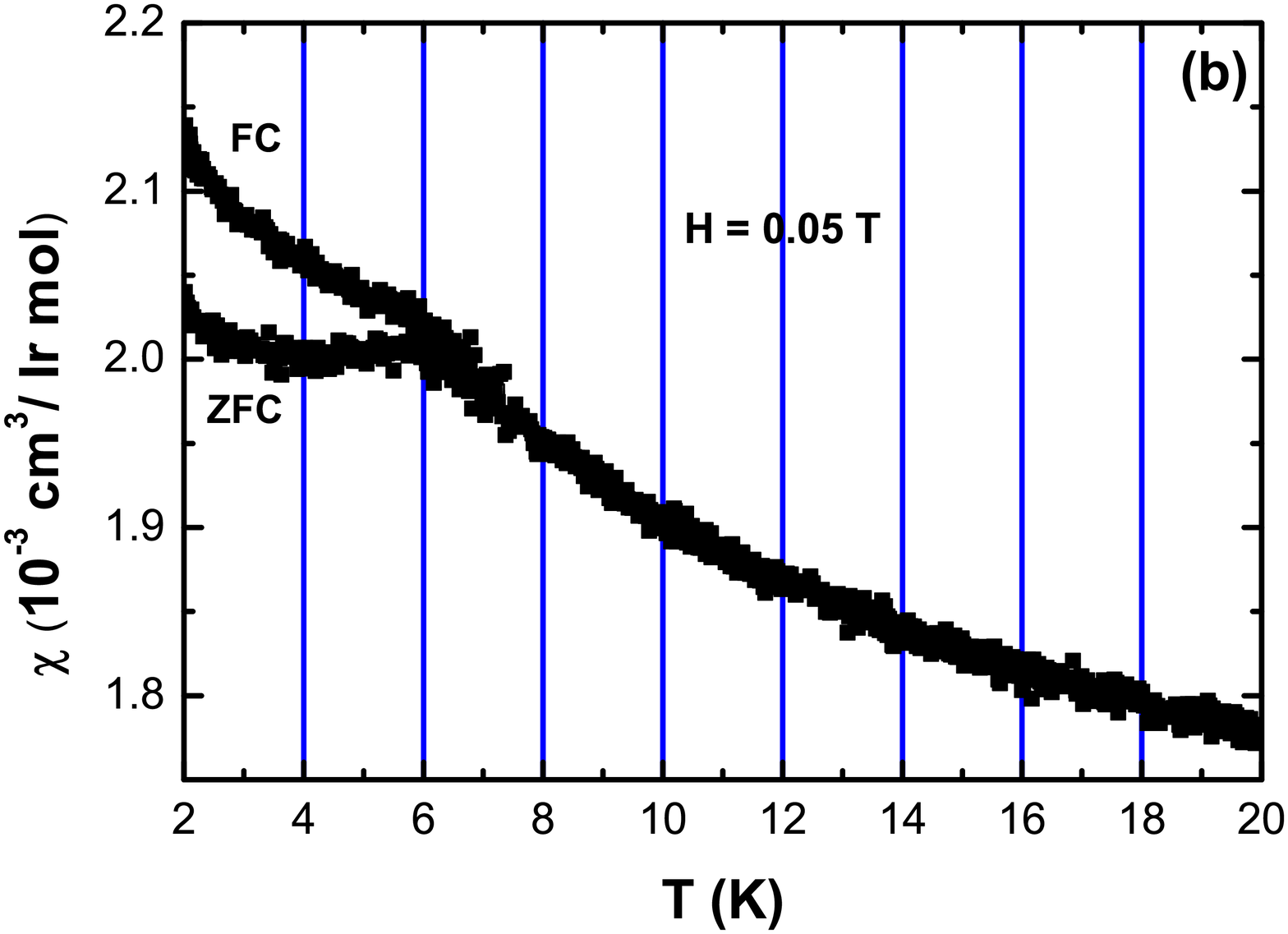}
\caption{(Color online) (a) Magnetic susceptibility $\chi$ versus $T$ of Na$_{4-x}$Ir$_3$O$_8$ ($x = 0, 0.1, 0.3, 0.7$) between $T = 2$~K and 400~K\@.  (b) Zero-field-cooled (ZFC) and field-cooled (FC) data measured in $H = 0.05$~T for the $x = 0$ parent compound.    
\label{Fig-chi}}
\end{figure}

Figure~\ref{Fig-chi}~(a) shows the magnetic susceptibility $\chi$ versus temperature $T$ between $T = 1.8$~K and $400$~K for Na$_{4-x}$Ir$_3$O$_8$ ($x \approx 0, 0.1, 0.3, 0.7$).  Figure~\ref{Fig-chi}~(b) shows the zero-field-cooled (ZFC) and field-cooled (FC) magnetic susceptibility $\chi$ versus temperature $T$ measured in a small magnetic field of $H = 0.05$~T for Na$_{4}$Ir$_3$O$_8$.  The bifurcation between the low field ZFC and FC data below $T \approx 6$~K suggests a frozen spin-glass like state develops below this temperature.  Similar low field behaviors were observed for Na$_{4-x}$Ir$_3$O$_8$ ($x \approx 0.1, 0.3$).  Observation of magnetic irreversibility below $T \approx 6$~K is consistent with previous reports \cite{Okamoto2007, Dally2014}.  

We now discuss the high field data shown in Figure~\ref{Fig-chi}~(a).  Data for two samples with $x \approx 0$ are shown and are marked with arrows in Fig.~\ref{Fig-chi}.  Although the temperature dependence of these two samples are similar, one can already see a slight difference in the absolute magnitude of $\chi(T)$ for two samples with apparently same Na content.  When Na is intentionally removed the magnetic susceptibility drops in magnitude although the local moment behavior persists up to at least $x \approx 0.3$.  The high temperature ($T \geq 200$~K) data were fit by the Curie-Weiss expression $\chi = \chi_0+{C\over T-\theta}$.  The parameters obtained from fits to the data for the samples Na$_{4-x}$Ir$_3$O$_8$ ($x = 0, 0.1, 0.3$) are given in Table~\ref{Table-chi}.  We see that the Curie constant $C$ monotonically decreases with increasing $x$ suggesting a decrease in the effective moment per Iridium in the Na deficient sample.  The value of the effective magnetic moments estimated from $C$ is also given in the Table~\ref{Table-chi}.  Additionally, the Weiss temperature $\theta$ decreases slightly but stays large and negative indicating that strong antiferomagnetic interactions persist between the surviving local magnetic moments.

\begin{table}

\caption{ Parameters obtained from fits to the magnetic susceptibility data by the Curie-Weiss expression $\chi = \chi_0 + {C \over T - \theta}$ }

\begin{ruledtabular}

\begin{tabular}{|c|cccc|}
x & $\chi_0$~($10^{-4}$~cm$^3$/Ir~mol) & $C$~(cm$^3$~K/Ir~mol) &  $\mu_{eff} (\mu_B)$&$\theta$~(K) \\ \hline  
0 & 1.2(1)& 0.39(1) & 1.77 &-568(9)  \\  
0.1 & 1.4(1)&0.33(2) & 1.62 &-512(6)  \\ 
0.3 & 1.8(6)&0.28(3) & 1.5 &-509(7) \\ 
0.7 & & & & \\ 
\end{tabular}

\end{ruledtabular}
\label{Table-chi}
\end{table} 

The $\chi(T)$ behavior for the $x = 0.7$ sample shows qualitatively different behavior.  The $\chi(T)$ is almost $T$ independent except the low temperature upturn which is seen for all samples.  The $T$ independent $\chi$ is similar to the behavior expected for a Pauli paramagnetic metal.  It must be noted however, that $\chi$ actually shows a weak increase with $T$ and thus is not the behavior of a simple metal.  This weak increase in $\chi(T)$ with increasing temperature is similar to what has been reported previously for Na$_3$Ir$_3$O$_8$~~\cite{Dally2014, Takayama2014}.  We believe, based on NMR evidence that our $x = 0.7$ sample is a distinct chemical phase.

\begin{figure}[t]
\includegraphics[width=3.25 in]{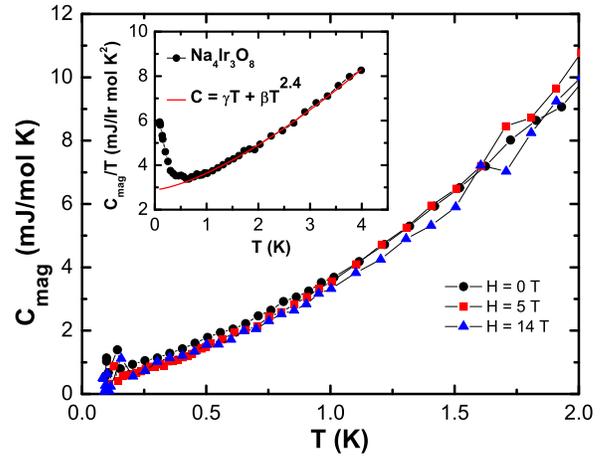}
\caption{(Color online) Low temperature magnetic heat capacity $C_{mag}$ versus temperature $T$ for the spin-liquid Na$_4$Ir$_3$O$_8$ between $T = 100$~mK and 2~K in magnetic fields $H = 0, 5, 14$~T\@.  The inset shows the $H = 0$ data plotted as $C_{mag}/T$ versus $T$.  The data above $T = 0.75$~K were fit (shown as the solid red curve through the data) by the expression $C = \gamma T + \beta T^p$ (see text for details).  
\label{Fig-C(T)}}
\end{figure}

\begin{figure}[t]
\includegraphics[width=3.25 in]{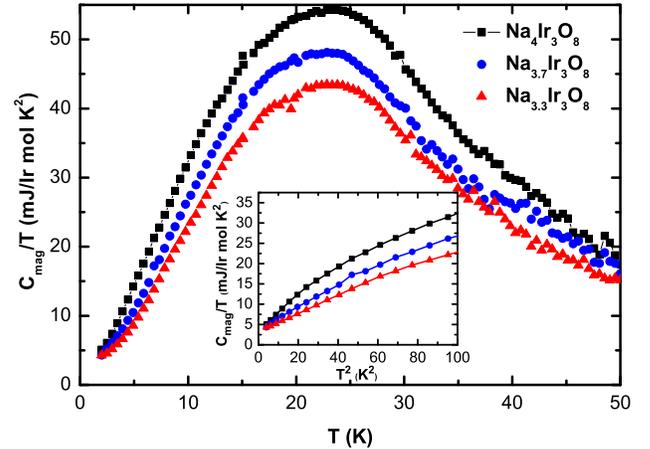}
\caption{(Color online) The magnetic contribution to the heat capacity $C_{mag}/T$ versus $T$ for Na$_{4-x}$Ir$_3$O$_8$ ($x = 0, 0.3, 0.7$).  (Inset) Low temperature $C_{mag}/T$ vs $T^2$ for Na$_{4-x}$Ir$_3$O$_8$ ($x = 0, 0.3, 0.7$).  
\label{Fig-Cmag(T)}}
\end{figure}

To confirm absence of any ordering, we have performed heat capacity measurements on Na$_4$Ir$_3$O$_8$ down to $T = 100$~mK\@.  In Fig.~\ref{Fig-C(T)} we show the low temperature magnetic heat capacity $C_{mag}$ versus $T$ for Na$_4$Ir$_3$O$_8$ measured at various magnetic fields $H = 0, 5, 14$~T between $T = 100$~mK and 2~K\@.  The magnetic contribution $C_{mag}$ was obtained by subtracting the $C(T)$ data of Na$_4$Sn$_3$O$_8$ (corrected for the molar mass difference) from the data of Na$_4$Ir$_3$O$_8$.  We do not see any evidence for magnetic ordering or freezing down to the lowest temperatures measured.  If the bulk of the sample goes into a frozen spin-glassy state, then one would expect to see a broad anomaly at a temperature of roughly 1.5--2 times the temperature at which the anomaly in the magnetic measurement is seen.  This would be around $T \approx 10$-- $12$~K for our samples with a magnetic anomaly at $T \approx 6$~K\@.  No such anomalies are observed in the heat capacity for any of our samples.  The $C_{mag}$ is also insensitive to applied magnetic fields up to $H = 14$~T as shown in Fig.~\ref{Fig-C(T)}.  The Fig.~\ref{Fig-C(T)} inset shows the zero field $C_{mag}/T$ vs $T$ between $T = 100$~mK and $T = 4$~K\@.  The data above $T = 0.75$~K were fit by the expression $C = \gamma T + \beta T^n$.  We were able to obtain an excellent fit, shown as the solid curve through the data in the inset of Fig.~\ref{Fig-C(T)}, with the values $\gamma = 2.8(4)$~mJ/K$^2$~mol~Ir, $\beta = 3.4(2)$~mJ/mol~K$^4$, and $n = 2.38(3)$.  The value of $\gamma \approx 2$~mJ/K$^2$~mol~Ir and an exponent $n$ between $2$ and $3$ is consistent with previous reports \cite{Okamoto2007, Singh2013}.

Below about 0.6~K we see an upturn in $C_{mag}/T$ which does not follow the $1/T^2$ behavior expected for the high temperature tail of a Schottky anomaly and is not field dependent.  A similar upturn was seen for $Na_3$Ir$_3$O$_8$ and was suggested to arise from an impurity contribution \cite{Fauque2014}.  The origin of this upturn for our sample is unknown at present. 

We now present the evolution with doping of another notable feature of the parent material: the broad anomaly in the magnetic contribution to the heat capacity at around $T \sim 30$~K [\onlinecite{Okamoto2007,Singh2013}].  Figure~\ref{Fig-Cmag(T)} shows the zero-field $C_{mag}/T$ vs $T$ for Na$_{4-x}$Ir$_3$O$_8$ ($x = 0, 0.3, 0.7$) between $T = 2$~K and $T = 50$~K\@.  The broad anomaly observed earlier for the undoped sample persists in the Na-deficient samples as well.  Although the magnitude of the anomaly decreases slightly, the maximum of the anomaly only reduces from $\approx 55$~mJ/mol~K$^2$ to about $\approx 42$~mJ/mol~K$^2$.  The position of the peak also doesn't change.  These observations suggest that the mechanism leading to the anomaly in Na$_4$Ir$_3$O$_8$ is also at work in the Na deficient samples.  A simpler and possibly disturbing possibility is that the anomaly in the magnetic heat capacity has nothing to do with the spin-liquid state but arises from an improper lattice subtraction.  This could happen if Na$_4$Sn$_3$O$_8$ does not have the same lattice heat capacity as Na$_4$Ir$_3$O$_8$.   

Figure~\ref{Fig-Cmag(T)}~inset, shows the zero-field $C_{mag}/T$ vs $T^2$ for Na$_{4-x}$Ir$_3$O$_8$ ($x = 0, 0.3, 0.7$) between $T = 2$~K and $T = 10$~K\@.  The downward curvature of the data for Na$_4$Ir$_3$O$_8$ in this plot highlights the power-law dependence of $C_{mag}(T)$ with an exponent between $2$ and $3$.  This exponent increases for the Na deficient samples with the data at the lowest temperatures ($T < 5$~K) for the $x = 0.7$ sample Na$_{3.3}$Ir$_3$O$_8$ showing a more conventional $T^3$ behavior.  It is also evident that $C_{mag}/T$ for all samples extrapolate to similar $T = 0$ values of $\gamma \approx 2.5$~mJ/Ir~mol~K$^2$.  

\emph{Summary and Discussion}: We have successfully synthesized Na deficient samples Na$_{4-x}$Ir$_3$O$_8$ ($x = 0, 0.1, 0.3, 0.7$) having the hyperkagome structure and performed electrical transport and thermodynamic measurements of the magnetic susceptibility $\chi(T)$, and heat capacity $C(T)$.  We have extended heat capacity measurements on the $x = 0$ parent compound down to $T = 100$~mK to look for any magnetic transition and have found none.  This firmly establishes Na$_4$Ir$_3$O$_8$ as a strong spin-liquid candidate material.  The parent compound also shows spin-freezing below $T \approx 6$~K in the magnetic measurements.  Microscopic measurements like NMR and muSR are desirable on our Na-deficient sample to check if bulk freezing is seen as has been observed recently \cite{Dally2014}.

For the Na deficient samples we find that the basic behavior of the parent $x = 0$ compound persists even on removing large amounts of Na.  Specifically, the local moment magnetism with large antiferromagnetic interactions is seen up to $x = 0.3$.  The broad anomaly in the magnetic heat capacity seen in the parent $x = 0$ material around $T = 30$~K is still seen even for the $x = 0.7$ sample although its magnitude decreases a bit.  Future experiments need to explore whether this anomaly arises from an improper lattice subtraction or is related to the materials magnetism.  We can also estimate the Wilson ratio $R_{\rm W}~=~{\pi^2 k_{\rm B}^2 \over 3\mu_{\rm B}^2}\bigg({\chi_P \over \gamma}\bigg)~.$ which is the ratio of the density of states probed by $\chi$ ($\chi_P$ is the Pauli paramagnetic susceptibility as $T \rightarrow 0$) to the density of states probed by heat capacity ($\gamma$ is the Sommerfeld coefficient) and is expected to be $1$ for a free-electron Fermi-gas.  For Na$_4$Ir$_3$O$_8$, using  $\chi_P = 9.4 \times 10^{-4}~{\rm cm^3/mol~Ir}$ and $\gamma = 2.5~{\rm mJ/mol~Ir}~{\rm K}^2$ from our recent measurements \cite{Singh2013} we find a quite large $R_W \approx 30$ suggesting strong magnetic correlations which enhance $\chi_P$ compared to $\gamma$.  Since the $\chi_P$ progressively decreases and $\gamma$ stays approximately the same as Na is removed, the $R_W$ progressively decreases for the Na deficient samples.  For the $x = 0.7$ sample, using $\chi_P \approx 2.2 \times 10^{-4}~{\rm cm^3/mol~Ir}$ and $\gamma$~=~$2.5~{\rm mJ/mol~Ir}~{\rm K}^2$ from Fig.~\ref{Fig-chi}~(a) and Fig.~\ref{Fig-Cmag(T)}~inset respectively, we obtain $R_W \approx 7$.  The reduced value of $R_W$ for the Na deficient samples is consistent with a recent theoretical work where it was shown that the susceptibility in the spin-liquid state is enhanced over the heat capacity due to strong spin-orbit coupling and correlations, which are reduced in the metallic state leading to a smaller $R_W$ for the metallic sample~ \cite{Chen2013}.  
 
Refinements of the powder diffraction patterns have shown that with increasing $x$, Na is progressively removed from the lattice.  Additionally we have found that Na is preferentially removed from the Na2 (4b) site.  This site is not on the Ir tetrahedra and hence deficiency in this site is not expected to lead to disorder in the magnetic Ir sub-lattice.  Therefore, the frustrated hyperkagome lattice of Ir moments is not disturbed with Na deficiency.  This is probably why the strong magnetic frustratuion is robust under removal of fairly large amounts of Na.  Na deficiency is however, expected to lead to hole doping of the system.  It is therefore surprising that local-moment behavior and insulating behavior survives under hole doping.  This is consistent with recent experiments on single crystals of Na$_{3+x}$Ir$_3$O$_8$ where it was found that for $x = 0.6$ (this would be $x = 0.4$ for our samples) the material was insulating~\cite{Fauque2014}.  A metallic state is obtained only for Na$_3$Ir$_3$O$_8$ where a different crystal structure is obtained \cite{Takayama2014}.  These observations are surprising given that Na$_4$Ir$_3$O$_8$ is regarded as sitting close to a metal-Insulator transition and near a quantum critical point (QCP) \cite{Singh2013}.          

In summary we have shown that the strongly frustrated Mott insulating state in Na$_4$Ir$_3$O$_8$ is quite robust against large removal of Na from the lattice.  The anomalous properties like the peak in $C_{mag}$ at $T \approx 30$~K, the power-law heat capacity, and the large Wilson ratio persist for the doped samples.  Any theory for the parent material should thus also be able to explain the above surprising behaviors.\\ \\ 
\noindent
Acknowledgments.-- We thank the X-ray facility at IISER Mohali for powder XRD measurements.  YS acknowledges DST, India for support through Ramanujan Grant \#SR/S2/RJN-76/2010 and through DST grant \#SB/S2/CMP-001/2013.

\end{document}